\pdfoutput=1
\RequirePackage[l2tabu, orthodox]{nag}
\newcommand*{\ATLASLATEXPATH}{}
\documentclass[UKenglish,texlive=2011,cernpreprint,PAPER]{\ATLASLATEXPATH atlasdoc}
\usepackage[full]{\ATLASLATEXPATH atlaspackage}
\usepackage{\ATLASLATEXPATH atlasbiblatex}
\usepackage{\ATLASLATEXPATH atlascontribute}
\usepackage{\ATLASLATEXPATH atlasphysics}
\graphicspath{{}}

\usepackage{multirow}
\usepackage{adjustbox}

\hypersetup{pdftitle={ATLAS draft},pdfauthor={The ATLAS Collaboration}}
\graphicspath{{logos/}}

\newcommand{\mcnlo}{\textsc{MC@NLO}}
\newcommand{\herwig}{\textsc{Herwig}}

\newcommand{\jimmy}{\textsc{Jimmy}}
\newcommand{\ctVER}[1][10]{\textsc{CT#1}}

\newcommand{\powheg}{\textsc{Powheg}}

\newcommand{\pythia}{\textsc{Pythia}}

\newcommand{\powhegpythia}{\powheg+\pythia}
\newcommand{\powhegherwig}{\powheg+\herwig}
\newcommand{\cteq}{\textsc{CTEQ6L1}}
\newcommand{\perugia}[1][2011c]{\textsc{Perugia#1}}

\usepackage{booktabs,siunitx}
\AtlasTitle{Measurement of colour flow with the jet pull angle in \ttbar events using the ATLAS detector at $\sqrt{s}=8$ TeV}

\AtlasJournal{Physics Letters B}

\AtlasAbstract{%
  The distribution and orientation of energy inside jets is predicted
  to be an experimental handle on
  colour connections between the hard--scatter quarks and gluons initiating
  the jets.  This Letter presents a measurement of the distribution
  of one such variable, the jet pull angle.
  The pull angle is measured for jets produced in \ttbar\ events with
  one \Wboson boson decaying leptonically and the other decaying to jets
  using 20.3~\ifb\ of data recorded with the ATLAS detector
  at a centre--of--mass energy of \rts~=~8~\TeV~at the LHC.
  The jet pull angle distribution is corrected for detector resolution
  and acceptance effects and is compared to various models.
}

\PreprintIdNumber{CERN-PH-EP-2015-128}

\newcommand{\allSMSigma}       {$0.8\sigma$}
\newcommand{\chargedSMSigma}       {$0.9\sigma$}

\newcommand{\allOctetSigma}       {$2.9\sigma$}
\newcommand{\chargedOctetSigma}       {$3.7\sigma$}

\begin{document}
\maketitle

\section{Introduction}
Due to the confining nature of the strong force, directly measuring the
quantum chromodynamic (QCD) interactions between quarks and gluons is not possible.
The strength and direction of the strong force depends on the colour
charge of the particles involved.
To a good approximation, the radiation pattern in QCD can be described through
a colour--connection picture, which consists of colour strings connecting quarks and gluons
of one colour to quarks and gluons of the corresponding anti--colour.
An important question is whether there is evidence of these colour connections
(\textit{colour flow}) in the observable objects: colour--neutral hadrons and the
jets they form.  The study of energy distributions inside and between jets in
various topologies has a long history, dating back to the discovery of gluons
in three--jet events at
PETRA~\cite{Brandelik:1979bd,Barber:1979yr,Berger:1979cj,Bartel:1979ut}.  Colour
connections are still a poorly constrained QCD effect,
which motivates the dedicated study presented in this Letter.
If well understood, experiments can exploit colour flow to aid Standard Model (SM)
measurements and searches for physics beyond the SM (BSM).
As a test that the colour flow can be extracted from the observable
final state, the data are compared to models with simulated $W$ bosons that are colour--charged or colour--neutral.

\begin{figure}[htbp]
  \centering
    \includegraphics[width=0.6\textwidth]{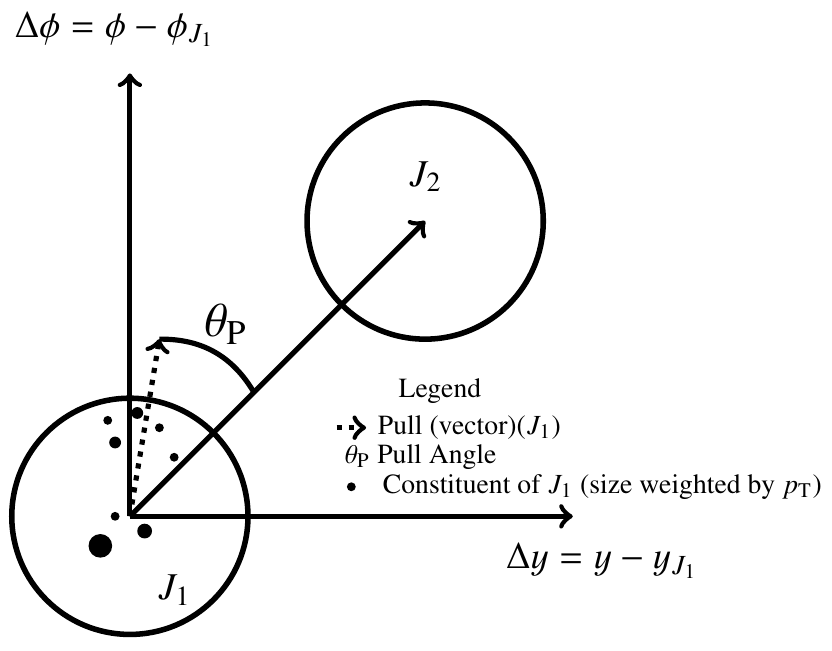}
  \caption{Diagram illustrating the construction of the jet pull angle for jet $J_1$ with respect to $J_2$.}
  \label{fig:pull:def}
\end{figure}

One observable predicted to contain information about the colour representation
of a dijet resonance like the $W$, $Z$,
or Higgs boson, is the \textit{jet pull vector}~\cite{Gallicchio:2010sw}.
The pull vector for a given jet $J$ with transverse momentum, $p_\text{T}^{J}$,
is defined as
\begin{equation}
  \vec{v}^{J}_{p} = \sum_{i\in J} \frac{p_\text{T}^i |\vec{r}_i|}{p_\text{T}^{J}}\vec{r}_i.
  \label{eq}
\end{equation}
The sum in Eq.~(\ref{eq}) runs over jet constituents with transverse momentum $p_\text{T}^i$ and location
$\vec{r}_i=(\Delta y_i,\Delta\phi_i)$, defined as the vector difference between the constituent and the
jet axis $(y_{J},\phi_{J})$ in rapidity ($y$) - azimuthal angle ($\phi$)
space\footnote{
  ATLAS uses a right--handed coordinate system with its
  origin at the nominal interaction point (IP) in the centre of the detector
  and the $z$--axis along the beam pipe.
  The $x$--axis points from the IP to the centre of the LHC ring,
  and the $y$--axis points upward. Cylindrical coordinates $(r,\phi)$ are used
  in the transverse plane, $\phi$ being the azimuthal angle around
  the beam pipe.
  The pseudorapidity is defined in terms of the polar angle
  $\theta$ as $\eta=-\ln\tan(\theta/2)$. Separation between objects in $(\eta, \phi)$ space is defined as $\Delta R = \sqrt{(\Delta\eta)^2 + (\Delta\phi)^2}$.  The rapidity of a four-vector $y=0.5\log\left(\frac{E+p_z}{E-p_z}\right)$, where $E$ is the energy and $p_z$ the component of the momentum parallel to the beam axis. }.
Given the pull vector for jet $J_1$, the angle formed between this pull vector and the vector connecting $J_1$ and another jet $J_2$,
$(y_{J_2}-y_{J_1},\phi_{J_2}-\phi_{J_1})$, is expected to be sensitive to
the underlying colour connections between the jets.
This is shown graphically in Fig.~\ref{fig:pull:def}, and the angle is called the
\textit{pull angle}, denoted $\theta_\text{P}(J_1,J_2)$.
The pull angle is symmetric around zero when it takes values between $-\pi$ and
$\pi$ and so henceforth $\theta_\text{P}(J_1,J_2)$ refers to the magnitude of
the angle in $(\Delta y,\Delta\phi)$ space with $0< \theta_\text{P}\leq \pi$.
If the pull vector is computed using jets originating from colour--connected
quarks, $\theta_\text{P}\sim 0$ since the radiation is predicted to fall
mostly between the two jets. If the vector is computed using two jets which
do not originate from colour--connected quarks,
there is no reason to expect $\theta_\text{P}$ to be small.
Thus $\theta_\text{P}$ should be useful for determining colour connections.

One of the challenges in studying colour flow is the selection of a final state
with a known colour composition.  Colour--singlet $W$ bosons from $t\bar{t}$
events provide an excellent testing ground because these bosons have a known
initial (colourless) state and such events can be selected with high purity.
The first study of colour flow using $W$ bosons from top quark decays was
performed by the D\O\ Collaboration~\cite{Abazov:2011vh}.
In the D\O\ analysis, calorimeter cells clustered within jets were used as
the constituents in Eq.~(\ref{eq}) and the resulting distribution was compared
to $W$ singlet (nominal) and $W$ octet templates.
The impact of the colour flow on the observable energy
distributions is very subtle; the D\O\ result was statistically limited
and was not able to significantly distinguish singlet
and octet radiation patterns.

The analysis presented in this Letter is a measurement of the jet pull angle in
$\sqrt{s}=8$ TeV $pp$ collisions at the Large Hadron Collider (LHC) with the ATLAS detector.
Instead of comparing the reconstructed jet pull angle in data directly
to simulated templates, as was done by the D\O Collaboration, the jet pull angle distribution
is first corrected for detector resolution and acceptance effects.
This allows for direct comparison with particle--level predictions for models
of physics beyond the SM as well as simulations of non--perturbative physics effects
with various tunable parameter values.

\section{Object and event selection}
\label{sec:events}

The ATLAS detector independently measures the inclusive and charged-particle energy
distributions in jets. This allows the jet pull angle to be constructed using
only the charged constituents of jets, or both the charged and neutral constituents.
In this analysis, both options are used in order to provide independent measurements
with different experimental systematic uncertainties.
Charged--particle momenta are measured in a series of tracking detectors
(collectively called the inner detector), covering a range of $|\eta| < 2.5$ and
immersed
in a 2 T magnetic field.
Electromagnetic and hadronic calorimeters surround the inner detector, with
forward calorimeters allowing electromagnetic and hadronic energy measurements
up to $|\eta| = 4.5$.
A detailed description of the ATLAS detector can be
found elsewhere~\cite{Detector}.

The anti--$k_t$ algorithm~\cite{Cacciari:2008gp}
with radius parameter $R=0.4$ is used to reconstruct jets from clusters of calorimeter
cells~\cite{TopoClusters} with deposited energy. The clusters are formed using
the local cluster weighting (LCW) algorithm
\cite{EndcapTBelectronPion2002} and calibrated to account for the detector
response as well as to mitigate the contributions from
additional $pp$ collisions per bunch crossing (pileup)~\cite{areasATLAS}.
The \textit{all--particles} pull angle is built from the clusters of calorimeter
cells assigned to a given jet.
In order to improve the rapidity resolution, jets and clusters are corrected
to point toward the reconstructed primary vertex\footnote{The primary vertex is
  defined as the vertex with the highest $\sum p_\mathrm{T}^2$ of associated tracks.}
  and the corrected jet four--vector
is used as the axis in Eq.~(\ref{eq}).

The \textit{charged--particles} pull angle is built from tracks that are associated
with a given jet~\cite{area,ATLAS:2012am}.
Tracks are reconstructed from hits in the inner detector and are required to
have $|\eta|<2.5$, $p_\text{T}>0.5$ \GeV, and satisfy various quality criteria,
such as association with the primary vertex, in order to suppress tracks from random hits and pileup tracks.
The charged--particles pull angle is constructed
using the four--momentum sum of all the associated tracks (treated as massless)
to provide the axis in Eq.~(\ref{eq}).  The {all--particles} and {charged--particles} pull angles are largely independent as they rely on information from different detector sub-systems.

In order to isolate a pure sample of hadronically decaying $W$ bosons,
this analysis targets a
$t\bar{t}\rightarrow b\bar{b} W(\rightarrow q\bar{q}^{\prime})W(\rightarrow \ell\nu)$
final state.

Events are selected by triggers requiring a single isolated electron
or muon, and the offline `tight' electron~\cite{Aad:2014fxa} or `combined'
muon~\cite{Aad:2014zya} must have $p_\text{T}>25$ \GeV\,
and $|\eta|<2.5$.
Basic quality criteria are imposed, including the existence
of at least one primary vertex associated with at least five tracks
with $p_\text{T}>0.4$ \GeV.
Furthermore, the magnitude $E_{\rm{T}}^{\mathrm{miss}}$
of the missing transverse momentum~\cite{Aad:2012re} is required to be greater than $20$ \GeV\,
and $E_{\rm{T}}^{\mathrm{miss}}+m_\text{T}>60$ \GeV\,, where $m_\text{T}$
is the transverse mass of the selected lepton and
the $E_{\rm{T}}^{\mathrm{miss}}$,
$m_\text{T} = \sqrt{2p_{\text{T}}^{\ell}E_{\rm{T}}^{\mathrm{miss}}(1-\cos(\phi^{\ell}-\phi^{\nu}))}$.
Events must have $\geq 4$ jets with $p_\text{T}>25$ \GeV\,.
At least two of these jets must be tagged as $b$--jets using the
multivariate discriminant, MV1 \cite{ATLAS-CONF-2012-043}, which uses impact parameter and
secondary vertex information.  The chosen MV1 working point corresponds to
an average $b$--tagging efficiency of 70\% for $b$--jets in simulated $t\bar{t}$
events.  At least two jets must not be $b$--tagged; of these,
the two leading--$p_\text{T}$ jets with $|\eta|<2.1$ are labelled as the jets from the
hadronically decaying $W$ boson, $J_i$ with $p_\text{T}^{J_1}>p_\text{T}^{J_2}$.
The $b$--tagged jets and the jets selected for the pull angle calculation
are required to have $|\eta|<2.1$
so that all constituents are within coverage of the inner detector used for
tracking.  
This procedure selects a sample that is expected to contain approximately
$90\%$ $t\bar{t}$ events. In 45\% of simulated $t\bar{t}$ events both jets
selected for the pull angle calculation contain energy from the
hadronically decaying $W$ boson.
Table~\ref{tab:comp} shows the predicted composition compared to the data yield.  More details about the various contributions are given below.

\begin{table}
  \centering
  \vspace{3mm}
  \begin{tabular}
  {
 c
 S[table-format=1,
   table-figures-uncertainty=0,group-digits = false,tight-spacing]
}
    \toprule
    Process                         & $\text{Number of events}$                  \\
    \midrule
    $t\bar{t}$                      & 95400 \hspace{1mm}$\pm$\hspace{1mm} 14000  \\
    $Wt$--channel single top         & 2730\hspace{1mm} $\pm$\hspace{1mm} 600     \\
    $s$-- and $t$--channel single top & 150\hspace{1mm} $\pm$\hspace{1mm} 10       \\
    $W$+jets                        & 3710\hspace{1mm} $\pm$ \hspace{1mm}120     \\
    $Z$+jets                        & 560 \hspace{1mm}$\pm$ \hspace{1mm}270      \\
    Dibosons                        & 190\hspace{1mm} $\pm$\hspace{1mm} 40       \\
    Multijets                       & 2500\hspace{1mm} $\pm$ \hspace{1mm}910     \\
    \midrule
    Total SM                        & 105000 \hspace{1mm}$\pm$ \hspace{1mm}14000 \\
    Data                            & 102987                                     \\
    \bottomrule
  \end{tabular}
  \caption{Estimated composition of the selected event sample.  The uncertainties are the sum in quadrature of the statistical uncertainties and either the uncertainties of the normalisation method (for the data driven W+jet and multi-jet estimates) or the uncertainties of the cross-section estimates. }
  \label{tab:comp}
\end{table}

\begin{figure}
  \centering
  \subfloat[]{
  \includegraphics[width=0.5\textwidth]{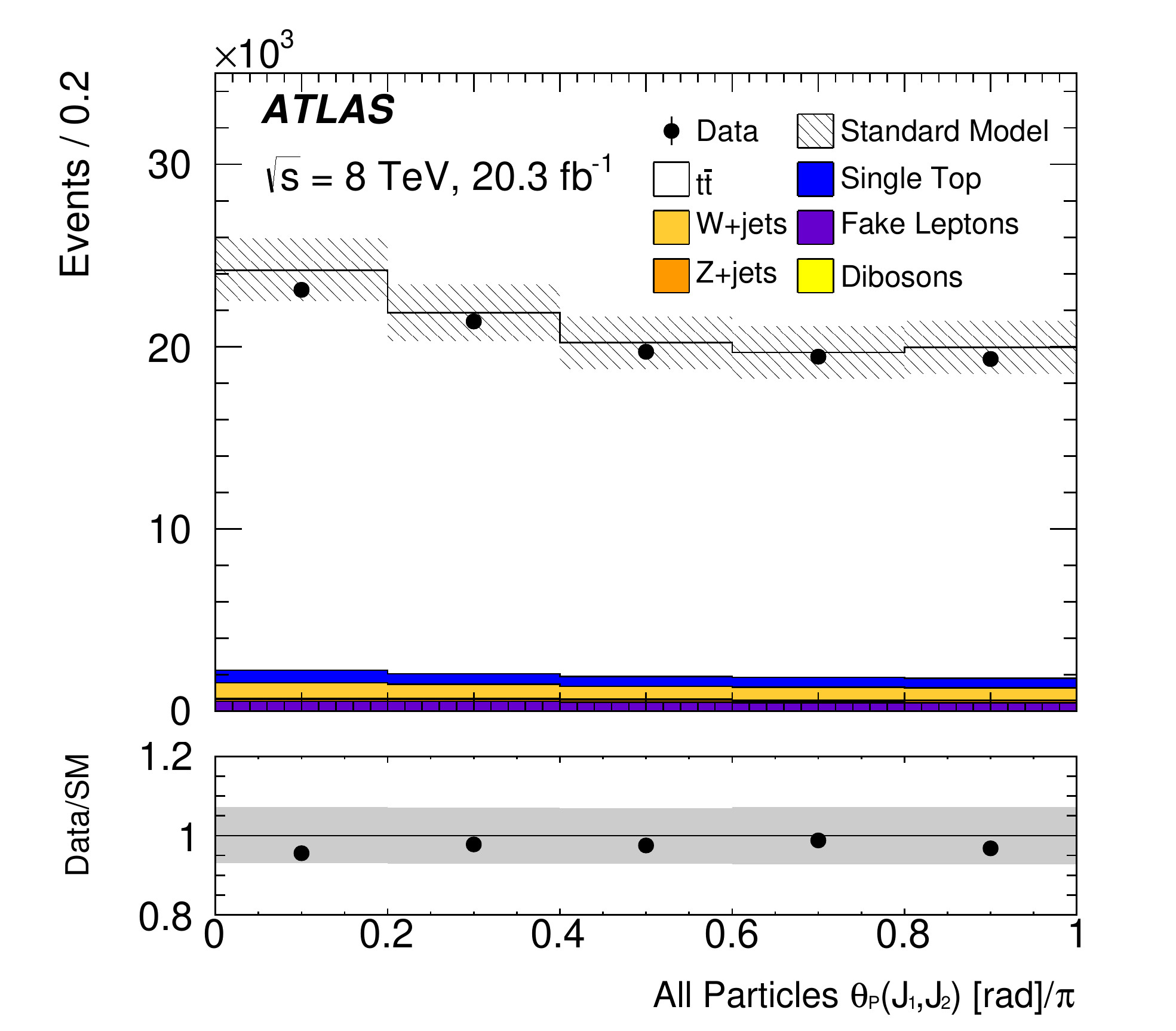}
  \label{fig:reco_all_particles_pull_angle}}
  \subfloat[]{
  \includegraphics[width=0.5\textwidth]{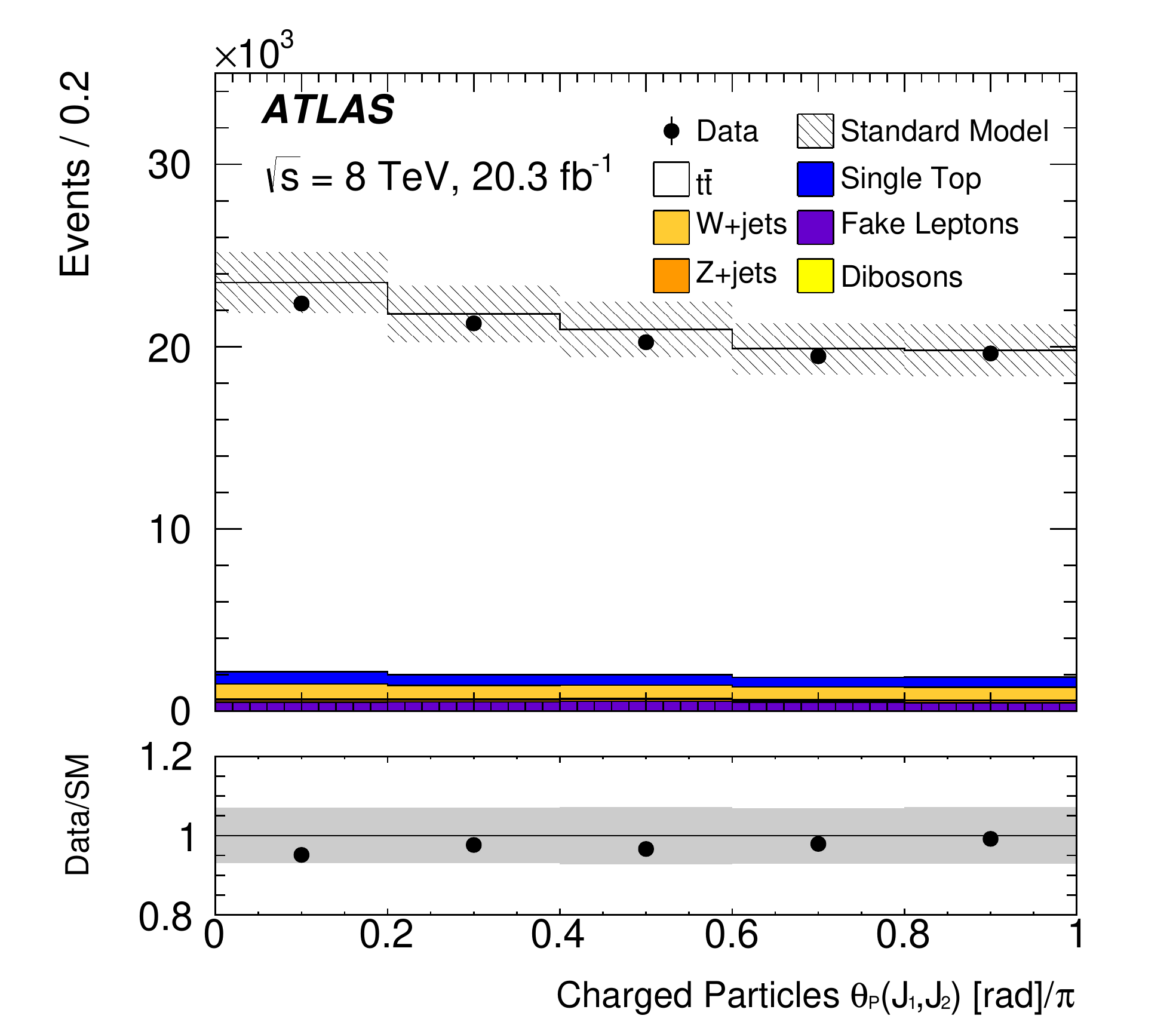}
  \label{fig:reco_charged_particles_pull_angle}}
  \caption{The detector--level \protect\subref{fig:reco_all_particles_pull_angle} all--particles
    and \protect\subref{fig:reco_charged_particles_pull_angle} charged--particles pull
    angle, $\theta_\mathrm{P}$, in data and in simulation.
    The uncertainty band includes only the experimental uncertainties on the
    inputs to the event selection and the jet pull calculation.
   A large part of the uncertainty displayed here affects the overall normalisation and is correlated between the individual bins.  This component of the uncertainty is cancelled in the unfolded measurement of the unit-normalised pull angle distribution.}
  \label{fig:reco_dists}
\end{figure}

There are two pull angles calculated per event, one calculated from the all--particles pull vector and one from the charged--particles pull vector of the highest $p_T$ jet assigned to the hadronic $W$ boson decay.  The all--particles and charged--particles pull angles are the angles that the corresponding pull vectors make with the direction from $J_1$ to $J_2$.   Figs. \ref{fig:reco_dists}\subref{fig:reco_all_particles_pull_angle}
and \ref{fig:reco_dists}\subref{fig:reco_charged_particles_pull_angle} show comparisons between data and simulation 
for the all-particles and charged-particles pull angles, calculated 
at detector level, i.e. from reconstructed objects.  Both distributions are broadly flat with an enhancement at small angles, consistent
with the SM prediction.

\section{Event simulation}
Monte Carlo (MC) samples are produced in order to determine how the detector response affects the pull angle
and to estimate some of the non--$t\bar{t}$ contributions in the data.
The details of the samples used are shown in Table \ref{tab:mc_samples}.

\begin{table}
  \centering
  \noindent\adjustbox{max width=\textwidth}{
  \begin{tabular}{cccccc}
    \toprule
    Process                               & Generator                                        & Type                 & Version & PDF                                  & Tune                          \\
    \midrule
                                                                                                                                                                                                     \\
    \multirow{2}{*}{\ttbar}               & \powheg~\cite{Nason:2004rx,Frixione:2007vw,pow3} & NLO ME               & -       & \ctVER~\cite{Lai:2010vv,Gao:2013xoa} & -                             \\
                                          & +\pythia~\cite{pythia6}                          & + PS                 & 6.426.2 & \cteq~\cite{Pumplin:2002vw}          & \perugia~\cite{Skands:2010ak} \\
                                                                                                                                                                                                     \\
    \multirow{2}{*}{Single top}           & \powheg                                          & NLO ME               &         & \ctVER(4f)                           & -                             \\
                                          & +\pythia                                         & + PS                 & 6.426.2 & \cteq                                & \perugia                      \\
                                                                                                                                                                                                     \\
    $WW,WZ,ZZ$                            & \textsc{Sherpa}~\cite{Gleisberg:2008ta}          & LO multi--leg ME + PS & 1.4.1   & \ctVER                               & Default                       \\
                                                                                                                                                                                                     \\
    \multirow{2}{*}{$W/Z$+jets}           & \textsc{Alpgen}~\cite{Mangano:2002ea}            & LO multi--leg ME      & 2.1.4   & \cteq                                & -                             \\
                                          & +\pythia                                         & + PS                 & 6.426.2 & \cteq                                & \perugia                      \\
                                                                                                                                                                                                     \\
    \midrule
                                                                                                                                                                                                     \\
    \multirow{3}{*}{\ttbar$^{\,\dagger}$} & \powheg                                          & NLO ME               & -       & \ctVER                               & -                             \\
                                          & +\herwig~\cite{Corcella:2000bw}                  & + PS                 & 6.520.2 & \ctVER                               & AUET2~\cite{AUET2}            \\
                                          & +\jimmy~\cite{JIMMY}                             & (MPI)                & 4.31    & -                                    & -                             \\
                                                                                                                                                                                                     \\
    \multirow{3}{*}{\ttbar$^{\,\dagger}$} & \mcnlo\cite{Frixione:2002ik,Frixione:2003ei}     & NLO ME               & 4.06    & \ctVER                               & -                             \\
                                          & +\herwig                                         & + PS                 & 6.520.2 & \ctVER                               & AUET2                         \\
                                          & +\jimmy                                          & (MPI)                & 4.31    & -                                    & -                             \\
                                                                                                                                                                                                     \\
    \bottomrule
  \end{tabular}}
  \caption{Monte Carlo samples used in this analysis. The abbreviations ME, PS, PDF, 
    MPI, LO and NLO respectively stand for matrix element, parton shower, parton distribution function,
    multiple parton interactions,
    leading order and next--to--leading order in QCD. Tune refers to the used set of 
tunable MC parameters.  Those samples marked
    with a $\dagger$ are used as alternative \ttbar\ samples to evaluate
    uncertainties due to the modelling of \ttbar\ events.  The nominal $t\bar{t}$ generator in the first line is used to estimate the yield in Table~\ref{tab:comp}.}
  \label{tab:mc_samples}
\end{table}

Aside from the $W$+jets background, all MC samples are normalised to
their theoretical cross--sections, calculated to at least next--to--leading order (NLO) precision in
QCD~\cite{Catani:2009sm,Kidonakis:2010tc,Kidonakis:2010ux,Kidonakis:2011wy,Campbell:1999ah,Campbell:2011bn}.
For the purpose of comparison between data and
the SM prediction before unfolding, \ttbar\ events are normalised to
a cross--section of $253\pm 15$ pb, calculated at
next--to-next--to--leading order (NNLO) in QCD
including next--to--next--to--leading logarithmic (NNLL)
soft gluon terms~\cite{Czakon:2011xx},
assuming a top--quark mass of $172.5$ \GeV.

Generated events are processed with a full ATLAS detector
and trigger simulation~\cite{ATLAS2010} based on
\textsc{Geant4}~\cite{Geant} and reconstructed using
the same software as the experimental data.
The effects of pileup are
modelled by adding to the generated hard--scatter events multiple minimum--bias
events simulated with \textsc{Pythia} 8.160 \cite{Sjostrand2008852},
the A2 set of tuned MC parameters (tune)~\cite{ATL-PHYS-PUB-2012-003} and the MSTW2008LO
Parton Distribution Function (PDF) set ~\cite{Martin:2009iq}. The distribution of the number of interactions is then weighted to
reflect the pileup distribution in the data.

\begin{figure}[htbp]
  \centering
    \includegraphics[width=0.6\textwidth]{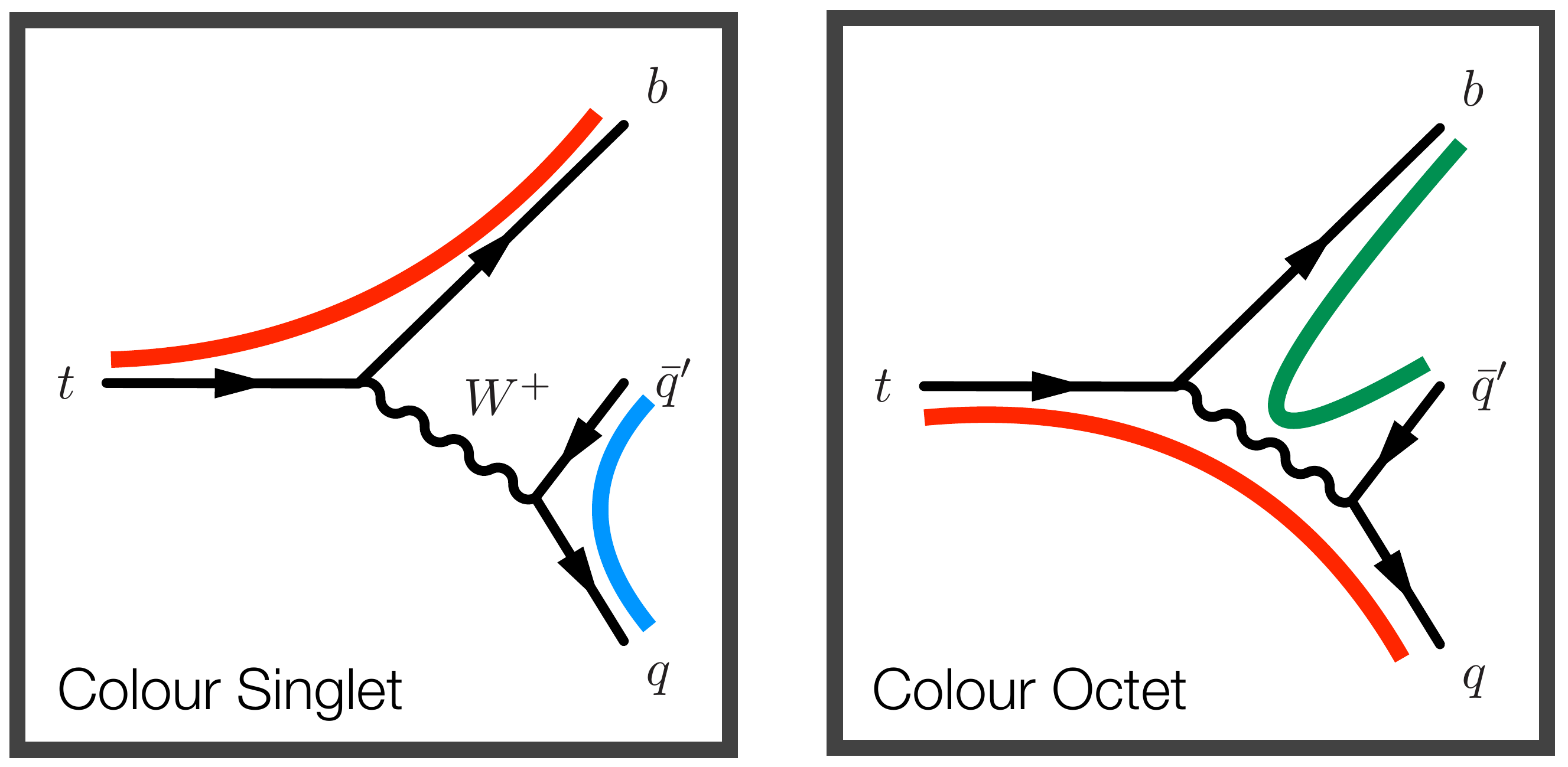}
  \caption{Diagram illustrating the colour connections (thick coloured 
lines) for the nominal sample with a colour-singlet $W$ (left) and the flipped sample with a colour-octet $W$ (right) samples in cases where the colour is distorted.}
  \label{fig:colourflow}
\end{figure}

To test the sensitivity of the jet pull angle to the singlet nature of
the $W$ boson, a sample was generated with a colour--octet $W$ boson.
The octet $W$ boson is simulated using the same setup as the nominal $t\bar{t}$
setup described in Table \ref{tab:mc_samples}.
Using the partons produced with \powheg\ (in the LHE~\cite{Alwall:2006yp}
format), the colour flow is inverted such that one of the $W$ decay daughters
shares a colour line with the $b$--quark and the other shares a line with the top quark,
as demonstrated schematically in Fig.~\ref{fig:colourflow}.
This sample is referred to as colour \textit{flipped} in the rest of this Letter.

\section{Unfolding}
\label{sec:unfolding}
In order to make direct comparisons with various theoretical models,
the data are unfolded to particle--level objects.
Particle--level jets are clustered from simulated particles with
a mean lifetime $\tau > 30$~ps,
before taking into account interactions with the detector.
The particle--level inputs to the all--particles pull angle are all of the
charged and neutral particles clustered within particle--level jets.
Only the charged particles clustered within the particle--level jets are used
for the charged--particles pull angle. Additional information
about the particle--level object definitions
can be found in Ref.~\cite{Aad:2015eia}.

The particle--level event selection is analogous to the detector--level
selection described in Sec.~\ref{sec:events} with detector--level
objects replaced with particle--level objects.
Exactly one electron or muon and at least four jets are required,
each with $p_\text{T} > 25$ GeV and $|\eta|<2.5$, with events discarded if
the electron or muon is within $\Delta R=0.4$ of a jet.
The particle--level
missing transverse momentum, $E_{\rm{T}}^{\mathrm{miss}}$,
calculated using the sum of neutrino four-momenta,
is required to be $E_{\rm{T}}^{\mathrm{miss}}>20$ GeV and the sum of
$E_{\rm{T}}^{\mathrm{miss}}+m_\text{T}>60$ GeV. At least two of the selected
jets are required to be identified as $b$--jets using the same definition
as that found in Ref.~\cite{Aad:2015eia}.
As with the detector--level calculation of the pull angle,
the two leading--$p_\text{T}$ particle--level non $b$--jets
with $|\eta|<2.1$ are labelled as the jets from the
hadronically decaying $W$ boson.

The first step of the unfolding procedure is to subtract from the data an estimate of
the non--$t\bar{t}$ backgrounds bin--by--bin in the pull angle distribution.
The $W$+jets and multijet backgrounds are estimated from the data using the
charge asymmetry and matrix methods, respectively~\cite{Aad:2014zka}.
The other backgrounds are estimated from simulation.
Single--top $Wt$ events have hadronically decaying $W$ bosons and are thus
sensitive to colour flow; however, even large changes in the colour flow through
the single--top contribution are subdominant compared to other uncertainties,
and are thus ignored.

After subtraction, the data are unfolded using an
iterative Bayesian (IB) technique~\cite{DAgostini10}
as implemented in the \textsc{RooUnfold} framework~\cite{Adye:2011gm}.
The IB method iteratively applies Bayes' theorem using the
response matrix to connect the prior to posterior at each step.
The response matrix is constructed from the nominal $t\bar{t}$ simulation
(also used for the prior) and describes the bin migrations between
the particle--level and detector--level pull angle distributions.
The binning of the response matrix and the number of iterations in the IB
method are chosen to minimise the overall uncertainty, described
in Section \ref{sec:systematics}.
In addition to correcting for bin migrations,
the unfolding procedure also corrects for events that pass either the
detector--level or particle--level selection, but not both.
The corrections are estimated using simulated \ttbar\ events.
Approximately 70\% of events that pass the detector--level selection also pass
the particle--level selection, while approximately 20\% of events that pass
the particle--level selection also pass the detector--level selection.
These corrections are found to be largely independent of the pull angle.

The all--particles pull angle has a stronger dependence on the colour flow,
but has a worse resolution than
the charged--particles pull angle\footnote{
The RMS of the all--particles
(charged--particles) pull angle is found to be $0.35\pi~(0.28\pi)$ radians.
} and is more
sensitive to the particle--level spectrum used as the IB prior.
Three bins and 15 iterations
are used for the all--particles pull angle and four bins
and three iterations are used for the charged--particles pull angle.
More iterations are required for the all--particles pull angle to reduce the
sensitivity to the IB prior, at the cost of a larger statistical uncertainty.
The size of each bin is comparable to the pull angle resolution.
The uncertainty on the unfolding procedure is determined in a data--driven
way by reweighting the particle--level distributions so that the corresponding detector--level distributions are in better agreement with the data~\cite{Malaescu}.  The method uncertainty is then estimated using the nominal response matrix by comparing the unfolded, reweighted simulation with the reweighted particle--level distribution.

\section{Systematic uncertainties}
\label{sec:systematics}
There are various systematic uncertainties which can impact
the measured jet pull angle.
The sources of uncertainty can be classified into two categories:
experimental uncertainties and theoretical modelling uncertainties. In the first category, some uncertainties impact the pull angle directly while the others impact only the acceptance.

The uncertainty in the energy scale and angular distribution of clusters of calorimeter cells are primary sources of uncertainty that directly impact the pull angle. To estimate the uncertainty in the angular resolution of clusters,
isolated tracks are matched to clusters in events enriched
in $Z(\rightarrow\mu\mu)$ + jets events.
The distribution of $\Delta R(\text{track},\text{cluster})$ is
investigated for various bins of track $p_\text{T}$ and $\eta$,
and in extrapolations to various layers of the calorimeter.
The observed resolution uncertainty is $\lesssim 1$~mrad,
but to account for the possible effects of multi--particle clusters, 5 mrad is
used, as in Ref.~\cite{daCosta:2011ni}.
Estimations of the cluster energy scale uncertainty are based on measurements
of the ratio of the cluster energy to the $p_\text{T}$ of matched tracks
($E/p$)~\cite{ATL-PHYS-PUB-2014-002}.
The ratio $E/p$ in 8 TeV data is compared with simulation and the differences
are used to estimate the uncertainty, which is parameterised as
$\alpha(1+\beta/p)$, where $\alpha$ and $\beta$ depend on $\eta$.
Typical values are $\alpha=0.05$ and $\beta=0.5$ \GeV\,.
To account for an uncertainty in the energy lost due to inactive material in the
detector and noise thresholds, low--energy clusters are randomly removed with
energy--dependent probabilities, as in Ref.~\cite{Aad:2012vm}.

For the charged--particles pull angle, there are uncertainties associated with
track reconstruction efficiency, which are estimated by randomly removing
tracks with an $\eta$--dependent probability~\cite{Aad:2010ac}.
Furthermore, tracks are randomly removed from jets with a
$p_\text{T}$--dependent factor to estimate the effect of uncertainties in
the efficiency of track reconstruction inside jets.
For jets with $p_\text{T}<500$ \GeV, the uncertainty on
the track reconstruction efficiency is less than $1\%$.

As the pull vector definition uses the calorimeter jet $p_\text{T}$,
both the all--particles and charged--particles pull angle are affected by
the uncertainty in the jet energy scale~\cite{Aad:2011he,Aad:2014bia}
and resolution~\cite{Aad:2012ag}.  However, changes in the jet energy scale and resolution do not impact the
pull \textit{angle}, but do impact the results via the acceptance due to
$p_\text{T}$ thresholds.
Similarly, uncertainties in the lepton energy scale, trigger efficiency,
$E_\text{T}^\text{miss}$ resolution and $b$--tagging
efficiencies~\cite{ATLAS-CONF-2012-040,ATLAS-CONF-2014-004,ATLAS-CONF-2012-043,ATLAS-CONF-2012-039}
indirectly affect the results through changes in acceptance.
Other sources of uncertainty on the acceptance, which impact the measurement
through the background subtraction, include those related to
the luminosity~\cite{Aad:2013ucp},
the multijet estimation,
and the normalisation and heavy flavour content of the $W$+jets
background~\cite{Aad:2014zka}.

The modelling of the $t\bar{t}$ sample is a primary source of theoretical uncertainty. For instance, there are different treatments of
the matrix element calculation (\powhegherwig\
versus \textsc{MC@NLO}+\textsc{Herwig})
and the parton shower model (\powhegpythia\
versus \powhegherwig).
The change in the result when using a response matrix constructed from
the flipped $t\bar{t}$ simulation is considered as a source of uncertainty
on the colour model.
Other sources of modelling uncertainties are evaluated,
including initial/final--state radiation by varying the
radiation simulated with \textsc{AcerMC} 3.8~\cite{Kersevan:2004yg} constrained
by Ref.~\cite{ATLAS:2012al}, multiple simultaneous parton collisions using the \textsc{Perugia} 2012 \textsc{mpiHi}
tune~\cite{Skands:2010ak}, colour reconnections with the \textsc{Perugia} 2012
\textsc{lowCR} tune, overlap between single top and $t\bar{t}$ by comparing
the diagram removal scheme with
the diagram subtraction scheme~\cite{Frixione:2008yi},
and PDF uncertainties~\cite{Botje:2011sn}.
\textit{Colour flow} describes the colour representation of the hard--scatter
process while \textit{colour reconnection}
is a phenomenological model for implementing a finite number of colours in the
parton shower.  The observable consequences of both processes are similar,
but colour reconnection has been constrained by 
charged--particle distributions in experimental data~\cite{Skands:2010ak}.
Varying the top quark mass by $\pm 1$ \GeV\ has a negligible impact on this measurement.

\begin{table}
  \centering
  \vspace{3mm}
  \begin{tabular}{ccc}
    \toprule
    Source              & All particles [\%] & Charged particles [\%] \\
    \midrule
    Unfolding procedure & 0.5                & 0.6                    \\
    Clusters            & 0.5                & N/A                    \\
    Tracks              & N/A                & 0.2                    \\
    JES                 & 0.4                & 0.2                    \\
    JER                 & 0.3                & 0.1                    \\
    Matrix element      & 1.5                & 0.9                    \\
    Shower model        & 1.7                & 0.6                    \\
    Colour model         & 1.3                & 1.0                    \\
    ISR/FSR             & 1.2                & 0.2                    \\
    MPI                 & 0.1                & 0.6                    \\
    Other               & 0.4                & 0.2                    \\
    \midrule
    Total systematic uncertainty    & 3.0                  & 1.8                    \\
    Statistical uncertainty        & 1.1                & 0.7                    \\
    \bottomrule
  \end{tabular}
  \caption{Summary of systematic uncertainties in the first bin of the all--particles and charged--particles pull angle distributions.  The jet energy scale, jet energy resolution, multiple simultaneous parton collision modeling, and initial/final--state radiation modeling uncertainties are abbreviated as JES, JER, MPI, and ISR/FSR, respectively.}

  \label{tab:systs}
\end{table}

The systematic uncertainties are estimated by unfolding the data with varied
response matrices or by
subtracting varied background predictions from the data.
The sources of uncertainty described above are summarised in
Table~\ref{tab:systs} for the first bin of the pull angle for both
the all--particles and charged--particles pull angle.
In general uncertainties are found to be larger for the
all-particles pull angle than the charged-particles pull angle
due to the remaining sensitivity to the IB prior, as discussed
in Section \ref{sec:unfolding}.
All bins carry information about the underlying colour flow,
but the first bin has the largest expected difference between the flipped
and nominal models.

\section{Results}

\begin{figure}[htbp]
  \centering
    \includegraphics[width=0.6\textwidth]{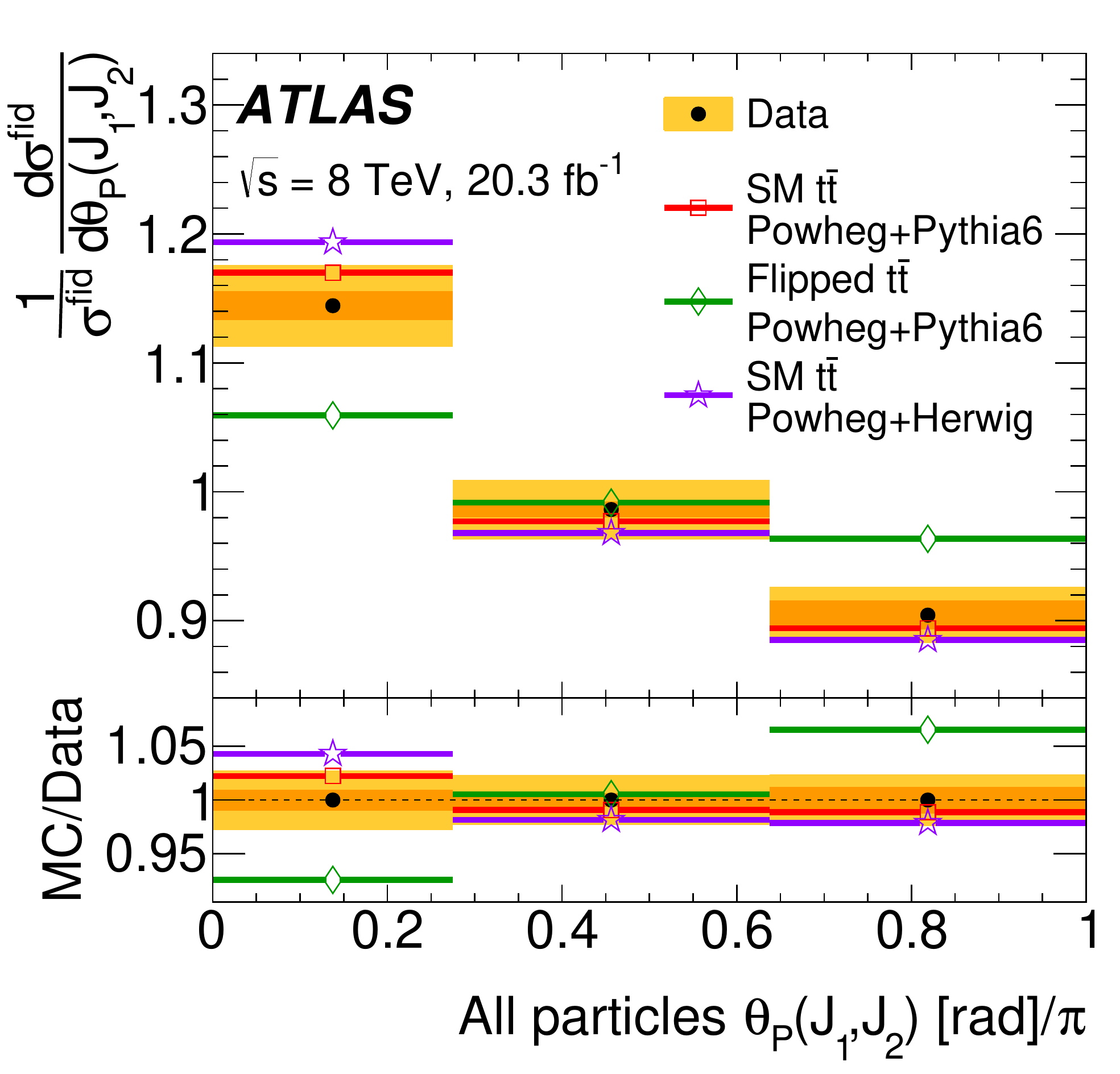}
    \caption{Normalised fiducial \ttbar\ differential cross--section for the jet pull angle distribution
      constructed using all particles.  The data are compared to two models with a colour-singlet 
$W$ boson (SM) but with different MC generators, and to the flipped model with 
a colour-octet $W$ boson. }
  \label{fig:unfolded:pull_all}
\end{figure}

\begin{figure}[htbp]
  \centering
    \includegraphics[width=0.6\textwidth]{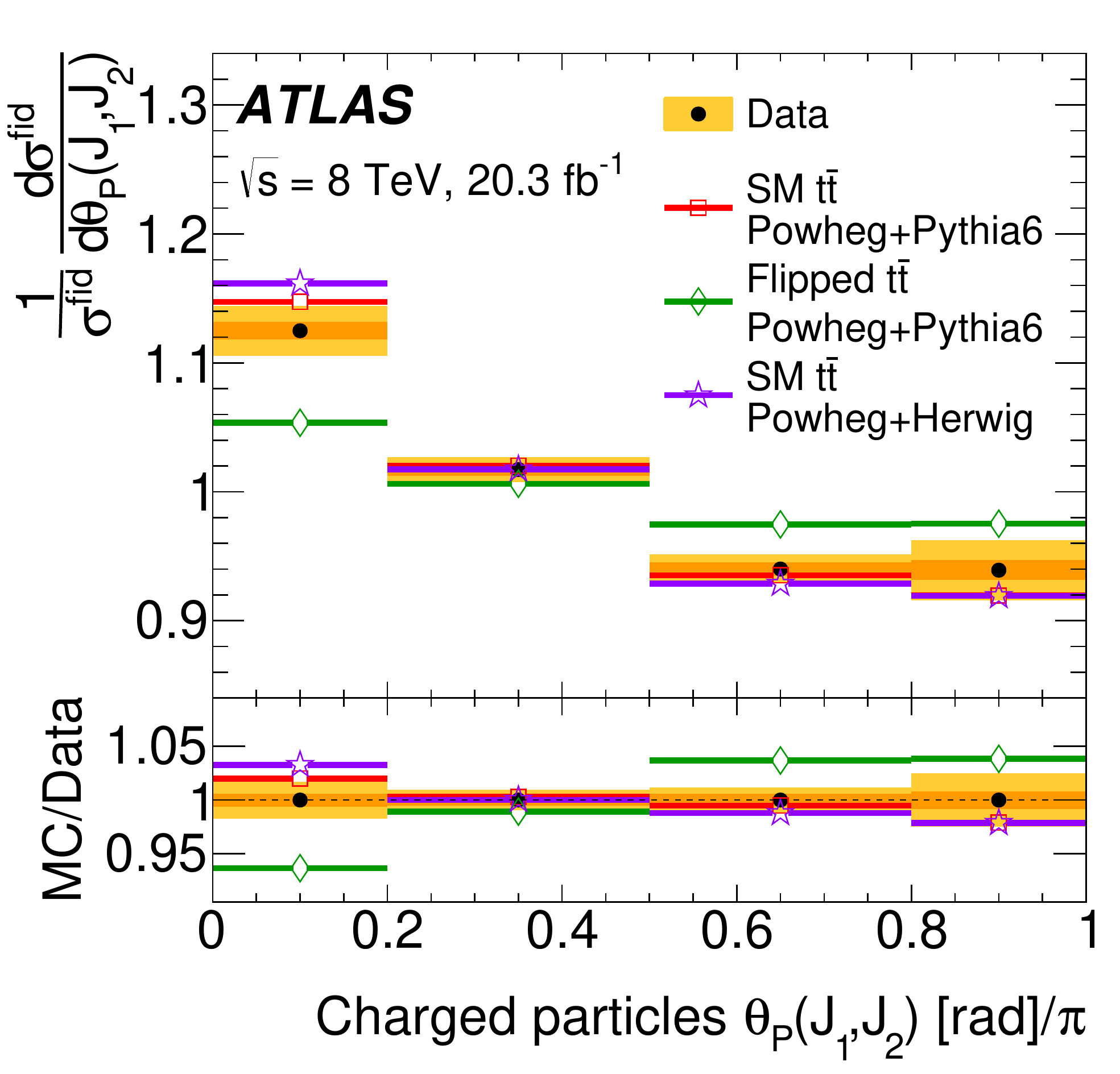}
  \caption{Normalised fiducial \ttbar\ differential cross--section for the jet pull angle distribution
      constructed using charged particles.  The data are compared to two models with a colour-singlet 
$W$ boson (SM) but with different MC generators, and to the flipped model with 
a colour-octet $W$ boson. }
  \label{fig:unfolded:pull_charged}
\end{figure}

The unfolded data are shown in Fig.~\ref{fig:unfolded:pull_all} (all--particles)
and Fig.~\ref{fig:unfolded:pull_charged} (charged--particles) and compared to
SM $t\bar{t}$ predictions simulated with \powhegpythia\
and \powhegherwig, and a flipped $t\bar{t}$ simulation generated using
\powhegpythia.

The data favour the SM predictions over the prediction of the flipped model.
Of the two SM predictions, \powhegpythia\ offers a slightly better description
of the data.
The level of agreement between the data and the \powhegpythia\
models is quantified using $\Delta\chi^2$ as a test statistic.
A $p$--value is computed taking
into account systematic and statistical uncertainties and their correlations by
generating pseudo--datasets from the full covariance matrix and then normalising the pull angle distributions.
This $p$--value is then converted into a one--sided Gaussian equivalent
$Z$--score.  The unfolded data are found to agree with
the nominal SM colour flow (with \powhegpythia)  at the \allSMSigma~(\chargedSMSigma) level
and differ from the flipped model at \allOctetSigma~(\chargedOctetSigma)
for the all--particles (charged--particles) pull angle.  Although the statistical
uncertainty is largely uncorrelated between the all--particles and
charged--particles pull angle measurements, the systematic uncertainty is not.
Since the charged--particles pull angle measurement is more sensitive than
the all--particles pull angle measurement, a full combination of the two
measurements does not provide increased sensitivity.

\section{Summary}

The analysis presented in this Letter describes a measurement of the orientation
of radiation from jets identified as originating from a $W$ boson in \ttbar\
events. The measurement uses 20.3 fb$^{-1}$ of $\sqrt{s}$ = 8 TeV $pp$ collision
data recorded by the ATLAS detector at the LHC.
To quantify the distribution of energy inside one jet relative to another,
the distribution of the \textit{jet pull angle} is extracted from the data using
information from both the ATLAS calorimeter and tracking detectors.
The jet pull angle is found to correctly characterise the $W$ boson as
a colour singlet, with data disfavouring an alternative colour--octet model
at greater than $3\sigma$.
This illustrates the potential to use the jet pull angle in future
SM measurements and BSM searches.
The jet pull angle measurement is presented as a normalised
fiducial \ttbar\ differential cross--section, allowing the results to be
used to constrain implementations of colour connection.

\section*{Acknowledgements}


We thank CERN for the very successful operation of the LHC, as well as the
support staff from our institutions without whom ATLAS could not be
operated efficiently.

We acknowledge the support of ANPCyT, Argentina; YerPhI, Armenia; ARC,
Australia; BMWFW and FWF, Austria; ANAS, Azerbaijan; SSTC, Belarus; CNPq and FAPESP,
Brazil; NSERC, NRC and CFI, Canada; CERN; CONICYT, Chile; CAS, MOST and NSFC,
China; COLCIENCIAS, Colombia; MSMT CR, MPO CR and VSC CR, Czech Republic;
DNRF, DNSRC and Lundbeck Foundation, Denmark; EPLANET, ERC and NSRF, European Union;
IN2P3-CNRS, CEA-DSM/IRFU, France; GNSF, Georgia; BMBF, DFG, HGF, MPG and AvH
Foundation, Germany; GSRT and NSRF, Greece; RGC, Hong Kong SAR, China; ISF, MINERVA, GIF, I-CORE and Benoziyo Center, Israel; INFN, Italy; MEXT and JSPS, Japan; CNRST, Morocco; FOM and NWO, Netherlands; BRF and RCN, Norway; MNiSW and NCN, Poland; GRICES and FCT, Portugal; MNE/IFA, Romania; MES of Russia and NRC KI, Russian Federation; JINR; MSTD,
Serbia; MSSR, Slovakia; ARRS and MIZ\v{S}, Slovenia; DST/NRF, South Africa;
MINECO, Spain; SRC and Wallenberg Foundation, Sweden; SER, SNSF and Cantons of
Bern and Geneva, Switzerland; NSC, Taiwan; TAEK, Turkey; STFC, the Royal
Society and Leverhulme Trust, United Kingdom; DOE and NSF, United States of
America.

The crucial computing support from all WLCG partners is acknowledged
gratefully, in particular from CERN and the ATLAS Tier-1 facilities at
TRIUMF (Canada), NDGF (Denmark, Norway, Sweden), CC-IN2P3 (France),
KIT/GridKA (Germany), INFN-CNAF (Italy), NL-T1 (Netherlands), PIC (Spain),
ASGC (Taiwan), RAL (UK) and BNL (USA) and in the Tier-2 facilities
worldwide.

\printbibliography

\clearpage
\newpage 
\begin{flushleft}
{\Large The ATLAS Collaboration}

\bigskip

G.~Aad$^{\rm 85}$,
B.~Abbott$^{\rm 113}$,
J.~Abdallah$^{\rm 151}$,
O.~Abdinov$^{\rm 11}$,
R.~Aben$^{\rm 107}$,
M.~Abolins$^{\rm 90}$,
O.S.~AbouZeid$^{\rm 158}$,
H.~Abramowicz$^{\rm 153}$,
H.~Abreu$^{\rm 152}$,
R.~Abreu$^{\rm 116}$
% ...
\bigskip
\\
% ...
$^{11}$ Institute of Physics, Azerbaijan Academy of Sciences, Baku, Azerbaijan\\
% ...
$^{85}$ CPPM, Aix-Marseille Universit{\'e} and CNRS/IN2P3, Marseille, France\\
% ...
$^{90}$ Department of Physics and Astronomy, Michigan State University, East Lansing MI, United States of America\\
% ...
$^{107}$ Nikhef National Institute for Subatomic Physics and University of Amsterdam, Amsterdam, Netherlands\\
% ...
$^{113}$ Homer L. Dodge Department of Physics and Astronomy, University of Oklahoma, Norman OK, United States of America\\
% ...
$^{116}$ Center for High Energy Physics, University of Oregon, Eugene OR, United States of America\\
% ...
$^{151}$ Institute of Physics, Academia Sinica, Taipei, Taiwan\\
$^{152}$ Department of Physics, Technion: Israel Institute of Technology, Haifa, Israel\\
$^{153}$ Raymond and Beverly Sackler School of Physics and Astronomy, Tel Aviv University, Tel Aviv, Israel\\
% ...
$^{158}$ Department of Physics, University of Toronto, Toronto ON, Canada\\
% ...
\end{flushleft}


\end{document}